\documentclass[useAMS, usenatbib, usegraphicx]{mn2e}
\usepackage{times}

\title[Radio emission from RS Oph in outburst]{Double radio peak and
  non--thermal collimated ejecta in RS~Ophiuchi following the 2006 outburst}

\author[S.~P.~S.~Eyres et al.]{S.~P.~S.~Eyres$^1$, 
T.~J.~O'Brien$^2$, R.~Beswick$^2$, T.~W.~B.~Muxlow$^2$, 
G.~C.~Anupama$^3$, 
\newauthor  N.~G.~Kantharia$^4$, M.~F.~Bode$^5$, M.~P.~Gawro{\'n}ski$^6$, R.~Feiler$^6$, A.~Evans$^7$, M.~T.~Rushton$^1$,
\newauthor R.~J.~Davis$^2$, T.~Prabhu$^3$, R.~Porcas$^8$, B. J. M. Hassall$^1$\\
$^1$Jeremiah Horrocks Institute for Astrophysics and Supercomputing, University of Central Lancashire,
Preston, PR1 2HE, UK\\
$^2$Jodrell Bank Centre for astrophysics, The Alan Turing Building, The University of Manchester, Manchester, M13 9PL, UK\\
$^3$ Indian Institute of Astrophysics, II Block Koramangala, Bangalore
560 034 India\\
$^4$ National Centre for Radio Astrophysics, Tata Institute of
Fundamental Research, Pune University Campus, Post Bag 3, Ganeshkind,
Pune 411007 India\\
$^5$Astrophysics Research Institute, Liverpool John Moores University,
Twelve Quays House, Egerton Wharf, Birkenhead, CH41 1LD, UK\\
$^6$Toru{\'n} Centre for Astronomy, Nicolas Copernicus University,
ul.~Gagarina~11, 87--100 Toru{\'n}, Poland\\
$^7$Astrophysics Group, Keele University,
Keele, Staffordshire, ST5 5BG, UK\\
$^8$Max--Planck--Institut f\"ur Radioastronomie, Auf dem H\"ugel 69, D--53121 Bonn, Germany\\}

\date{Accepted 2009 February 12.  Received 2009 February 10; in original form 2008 October 6}

\begin{document}
\label{firstpage}
\maketitle

\begin{abstract}
We report MERLIN, VLA, OCRA-p, VLBA, Effelsberg and GMRT observations
beginning 4.5~days after the discovery of RS~Ophiuchi undergoing its
2006 recurrent nova outburst. Observations over the first 9~weeks are
included, enabling us to follow spectral development throughout the
three phases of the remnant development. We see dramatic brightening
on days~4 to 7 at 6~GHz and an accompanying increase in other bands,
particularly 1.46~GHz, consistent with transition from the initial
``free expansion'' phase to the adiabatic expansion phase. This is
complete by day~13 when the flux density at 5~GHz is apparently
declining from an unexpectedly early maximum (compared with
expectations from observations of the 1985 outburst). The flux density
recovered to a second peak by approximately day~40, consistent with
behaviour observed in 1985. At all times the spectral index is
consistent with mixed non--thermal and thermal emission. The spectral
indices are consistent with a non--thermal component at lower
frequencies on all dates, and the spectral index changes show that the
two components are clearly variable.
The estimated extent of the emission at 22~GHz on day~59 is consistent
with the extended east and west features seen at 1.7~GHz with the VLBA
on day~63 being entirely non--thermal. We suggest a two--component
model, consisting of a decelerating shell seen in mixed thermal and
non-thermal emission plus faster bipolar ejecta generating the
non--thermal emission, as seen in contemporaneous VLBA observations.
Our estimated ejecta mass of 4$\pm$2$\times10^{-7}$~M$_\odot$ is
consistent with a WD mass of 1.4~M$_\odot$. It may be that this ejecta
mass estimate is a lower limit, in which case a lower WD mass would be
consistent with the data.
\end{abstract}


\section{Introduction}
\label{sec-intro}

RS~Ophiuchi is the most active and best studied member of the small
Recurrent Nova class of interacting binary star. It consists of a red
giant (RG) donating material to a white dwarf (WD) primary, via either
Roche Lobe overflow or direct wind accretion.  Six optical outbursts
have been observed at irregular intervals (1898, 1933, 1958, 1967,
1985 and 2006) with two others suggested in the literature
\citep[1907; 1945
from][respectively]{Schaefer2004,Oppenheimer1993}. The optical
development is very similar in each case \citep{Rosino1986}. The 1985
outburst was the first in which observations were made at radio
wavelengths, with first detection 18~days after outburst
\citep{Padin1985}.  Spectral development showed both non-thermal
synchrotron and thermal free-free emission, varying over the course of
the rise to peak by about day~40 and subsequent decline over 240~days
\citep{Hjellming1986}. A single VLBI image was obtained, showing
possible three--component emission extended east--west around the
radio peak \citep{Taylor1989}.

%

The latest outburst by this object was discovered on 2006~February~12,
and reached a peak of V=4.5 on 2006~February~12.83 \citep{Narumi2006},
which we take as day~0 for the rest of this paper. We began radio
observations on day~4, first with MERLIN at 6~GHz and then with the
VLA at 1.49, 4.89, 14.96 and 22.48~GHz (L, C, U and K bands
respectively), as described in this paper. In addition VLBI
observations began on day~13 and continued throughout the period of
the observations discussed here. Both the VLBI \citep{OBrien2006} and
X-ray \citep{Bode2006} observations are consistent with an expanding
shock wave due to the fast ejecta from the eruption on the WD
encountering the pre--outburst RG wind. HST observations on day~155
\citep{Bode2007} gave added weight to the suggestion of a bipolar
remnant from VLBI observations by \citet{OBrien2006}. Here we
interpret the first and second radio peaks with reference to the VLBI,
HST and X-ray observations



\section{Observations}
\label{sec-observations}

\subsection{MERLIN}
\label{ssec-MERLIN_obs}

The Multi-Element Radio-Linked Interferometer-Network (MERLIN)
responded rapidly to the triggering of target-of-opportunity
observations, with the first observations starting at
day~4.3. Monitoring continued at times complementary to VLA, the
European VLBI Network (EVN) and the Very Long Baseline Array (VLBA),
with observation dates as shown in Table~\ref{tab-obs}. Observations
were made initially at 5~cm and later at 6~cm (collectively referred
to as C~band). Full imaging runs were made, at around 10~hours each
including calibration observations. The flux calibrator in each case
was 3C286, with OQ208 used for point source response and polarisation
calibration.  Phase calibration was performed with regular pointings
at J1743$-$038. 
Using standard values and models for the flux calibrator,
the flux scale for the phase calibrator was determined. The phase
calibrator solutions were then interpolated to the RS~Oph data. For
the observations on days~4 and 5, the flux density was rising so rapidly that
the data were divided into 30--min segments and we were able to
follow the rapid rise. A similar analysis on day~7 showed that the
rate of decline had flattened by that date.

\subsection{VLA}
\label{ssec-VLA_obs}

The Very Large Array (VLA) in New Mexico also responded rapidly to the
request for observations, with the first made on 2006~February~17
(day~4.7), almost simultaneous with the first MERLIN observations, and
continuing approximately weekly thereafter. Observation dates are
given in Table~\ref{tab-obs}, and on each occasion continuum
observations were made at 1.46~GHz (L~band), 4.89~GHz (C~band),
14.96~GHz (U~band) and 22.48~GHz (K~band) with a 50~MHz bandwidth.
Observations were made in standard interferometer mode at A~array (the
most extended and high--resolution configuration), with 1743$-$038
(J2000) as secondary calibrator. Either 3C286 or 3C48 was used as the
primary calibrator, depending upon which was available within the
scheduled observing window. 
Fluxes were calculated for the primary calibrators using the approach
outlined in \cite{Baars1977} but employing the latest (1999.2) VLA
coefficients. The phase and amplitude solutions were determined using
model images supplied within AIPS. The secondary calibrator solutions
were determined assuming it to be a point source, and the flux at
observation from these solutions and the calculated flux of the
primary calibrator. Finally the flux scale for RS~Oph was determined
from the secondary calibrator solutions. Images were generated using
the CLEAN algorithm and AIPS task IMAGR at each band on each date and
used to determine flux densities. At 22.48~GHz it was also feasible to
estimate the extent of the emission as these were the data with the
highest spatial resolution.



\subsection{OCRA-p}
\label{ssec-OCRA_obs}

30-GHz continuum observations were made using the Toru{\'n} 32-m radio
telescope and a prototype (two-element receiver) of the One-Centimeter
Radio Array \citep[OCRA-p,][]{Lowe2005}. The recorded output from the
instrument was the difference between signals from two closely-spaced
horns effectively separated in azimuth by 3.1 arcminutes so the
atmospheric and gain fluctuations were mostly cancelled out. The
observing technique was such that the respective two beams were
pointed at the source alternately with a switching cycle of $\sim$50~s
for a period of $\sim$5.5~min, thus measuring the source flux density
relative to the sky background on either side of the source. The
primary flux calibrator was the planetary nebula NGC~7027, which has
an effective radio angular size of 8~arcsec
\citep{Bryce1997}. Assuming a spectral index of $-0.1\pm0.1$ and using
an absolute temperature calibration of NGC~7027 \citep[giving
5.45$\pm$0.20~Jy at 32~GHz, from][]{Mason1999} we estimate the flux
density of NGC~7027 at 30~GHz as 5.46$\pm$0.20~Jy. All the final flux
density values have been scaled to this figure. The telescope pointing
was determined from the azimuth and elevation scans across the nearby
point source J1743$-$038. The same source was also used as a secondary
flux density calibrator because NGC~7027 was at some distance from the
target source. Corrections for the effects of the atmosphere were
determined from the system temperature measurements at zenith distances
of 0$^\circ$ and 60$^\circ$. 


\subsection{VLBA}

Observations were made at 1.667~GHz with the Very Long Baseline Array
(VLBA) on day~63 (2007~April~17) -- see \citet{OBrien2008} for more
details. The data were correlated at the VLBA Correlator, Array
Operations Center, New Mexico before being transferred to Jodrell Bank
Observatory, United Kingdom and Joint Institute for VLBI in Europe in
Dwingeloo, Netherlands for calibration and mapping. Primary flux
calibration was carried out using 3C286, with secondary calibration on
J1745-0753. As the resolution at this epoch was around 20~mas it was
necessary to model the structure of both calibrators to obtain
reliable flux and phase calibration. Fringe fitting was carried out to
complete calibration and allow imaging of the emission structure.




\section{Results}
\label{sec-results}

The flux density values from each date at each observed frequency are given in
Table~\ref{tab-obs}.  In addition Effelsberg measured 36 mJy at 10.45
GHz on day~5.5. We refer to \citet{Kantharia2007} for Giant Metrewave
Radio Telescope (GMRT) values below 1.4~GHz.

\begin{table*}
\begin{tabular}{llllcccccc}

Date	& JD & Day & Telescope & \multicolumn{6}{c}{Frequency (GHz)} \\
&  & & & 30 & 22.48 & 14.96 &  6 & 4.89  & 1.46  \\
\hline
17/2	 & 783.83 & 4.50$^*$  & M   & - & - & - & 14$\pm$2 &  - & - \\
17/2	 & 784.02 & 4.70  & V   & - & 26.2$\pm$0.5 & 23.2$\pm$0.6  & -
& 15.2$\pm$0.2 & 2.8$\pm$0.2 \\
18/2	 & 784.82 & 5.50$^*$  & M   & - & - & - & 33$\pm$2& - & - \\
20/2	 & 786.82 & 7.46  & M   & - & - & - & 41.2$\pm$0.8  & -  & - \\
26/2	 & 793.21 & 13.88 & V   & - & 61.1$\pm$0.7 &  50.5$\pm$0.5& - & 53.3$\pm$0.2& 57.6$\pm$0.3 \\
\hline
01/3	 & 795.66 & 16.34 & O   & 55$\pm$6    & - & - & - & - & - \\
01/3	 & 795.62 & 16.52 & M   & - & - & - &  37.9$\pm$1.1& - & - \\
02/3	 & 796.62 & 17.47 & M   & - & - & - &  35.8$\pm$0.7& - & -  \\
06/3	 & 801.18 & 21.85 & M, V   & - & 36.4$\pm$0.9& 37.4$\pm$0.6&  43$\pm$5&44.5$\pm$0.2 &48.2$\pm$0.3\\
08/3	 & 802.68 & 23.36 & O   & 53$\pm$6   & - & - & - & -  & - \\
09/3	 & 803.61 & 24.28 & O   & 74$\pm$7    & - & - & - & -  & - \\
13/3	 & 807.18 & 27.85 & V   & - & 69.4$\pm$0.7 & 52.9$\pm$0.6& - &50.3$\pm$0.2 & 50.4$\pm$0.4\\
14/3	 & 807.83 & 28.50 & M & - & - & - & 45$\pm$5 & - & -  \\
23/3	 & 817.73 & 38.41 & O   & 103$\pm$8   & - & - & - & - & - \\
23/3	 & 818.95 & 39.62 & V& - & 78.1$\pm$0.5& 49.9$\pm$0.7& - &51.2$\pm$0.2 &55.5$\pm$0.3 \\
24/3	 & 819.95 & 40.62 & M   & - & - & - & 50$\pm$5 & - &  - \\
25/3	 & 820.95 & 41.62 & V   & - & 78.5$\pm$0.8& 55.8$\pm$0.7& - &51.7$\pm$0.2 &53.8$\pm$0.3\\
30/3	 & 825.90 & 46.57 & V   & - & 65.4$\pm$0.6& 48.8$\pm$0.6& - & 47.4$\pm$0.2 &52.9$\pm$0.2\\
\hline
03/4	 & 829.04 & 49.71 & M   & - & - & - & 40$\pm$5 & - & - \\
06/4	 & 832.04 & 52.71 & V   & - & 66.0$\pm$0.1    & 50.9$\pm$0.7& - &42.4$\pm$0.2&48.6$\pm$0.3\\
13/4	 & 839.07 & 59.74 & V   & - & 79.9$\pm$1.7 & 50.5$\pm$0.7& - &39.3$\pm$0.2 &42.8$\pm$0.3\\
\end{tabular}
\caption{Flux density values (mJy) from observations with MERLIN (M),
  the VLA (V), OCRA-p (O) and GMRT (G). No data were taken on dates
  marked -.  Uncertainties on dates marked $^*$ account for the
  increasing flux density over the period of each observation.  Date
  is day/month in 2006. JD is the Julian Date relative to
  JD2453000. Day is number of days after optical discovery (centre of
  observing period).}



\label{tab-obs}
\end{table*}


In order to follow the developments more clearly, we have plotted
radio light curves and spectral energy distributions.

\subsection{Light curves}
\label{ssec-lightcurves}

\begin{figure*}
\includegraphics[angle=270, width=18.5cm]{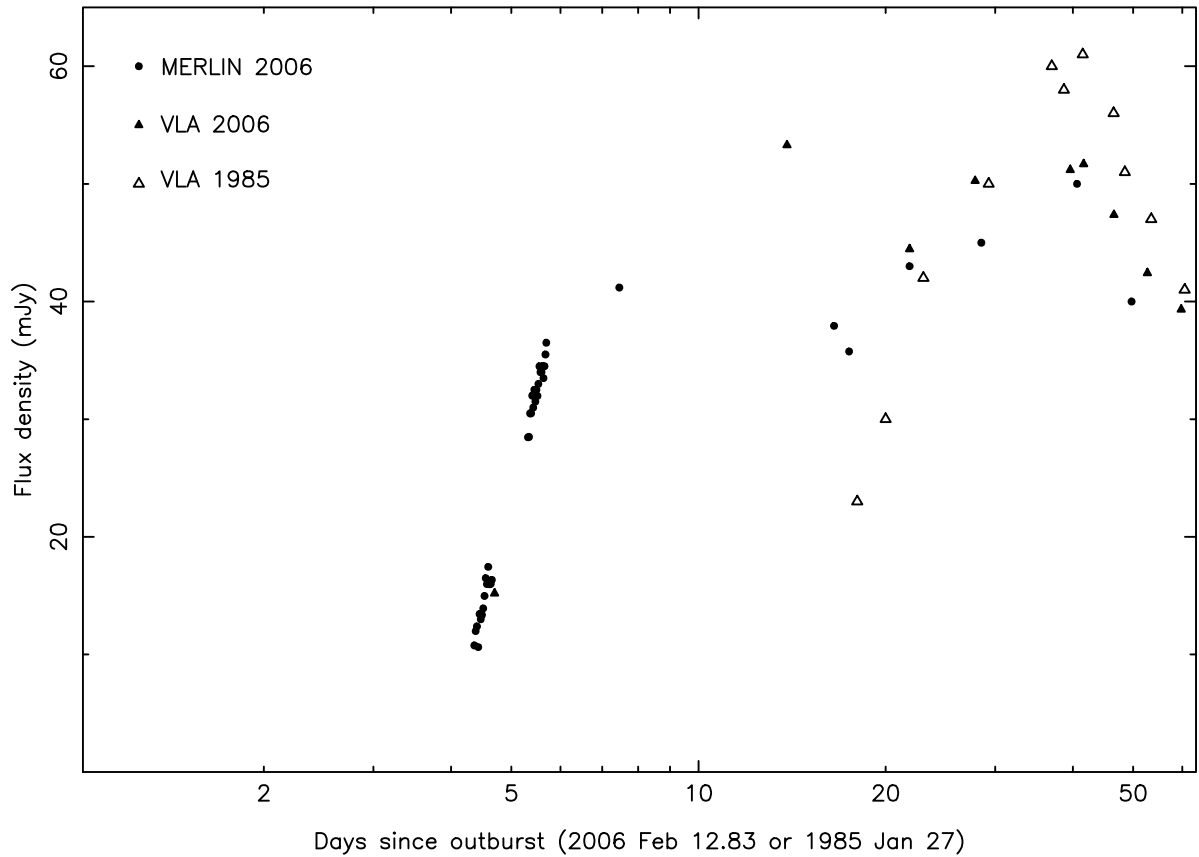}
\caption{Radio lightcurves at 5 to 6~GHz. Symbols: 2006 VLA=filled
  triangles; MERLIN=filled circles (small ones during the initial rise
  -- error bars not included); 1985 VLA open triangles -- observations
  during the 1985 outburst from \citet{Hjellming1986}. Uncertainties
  are smaller than or comparable to the symbol size except where error
  bars indicate otherwise.}
\label{fig-Cband}
\end{figure*}

\begin{figure}
\includegraphics[angle=0, height=20.5cm]{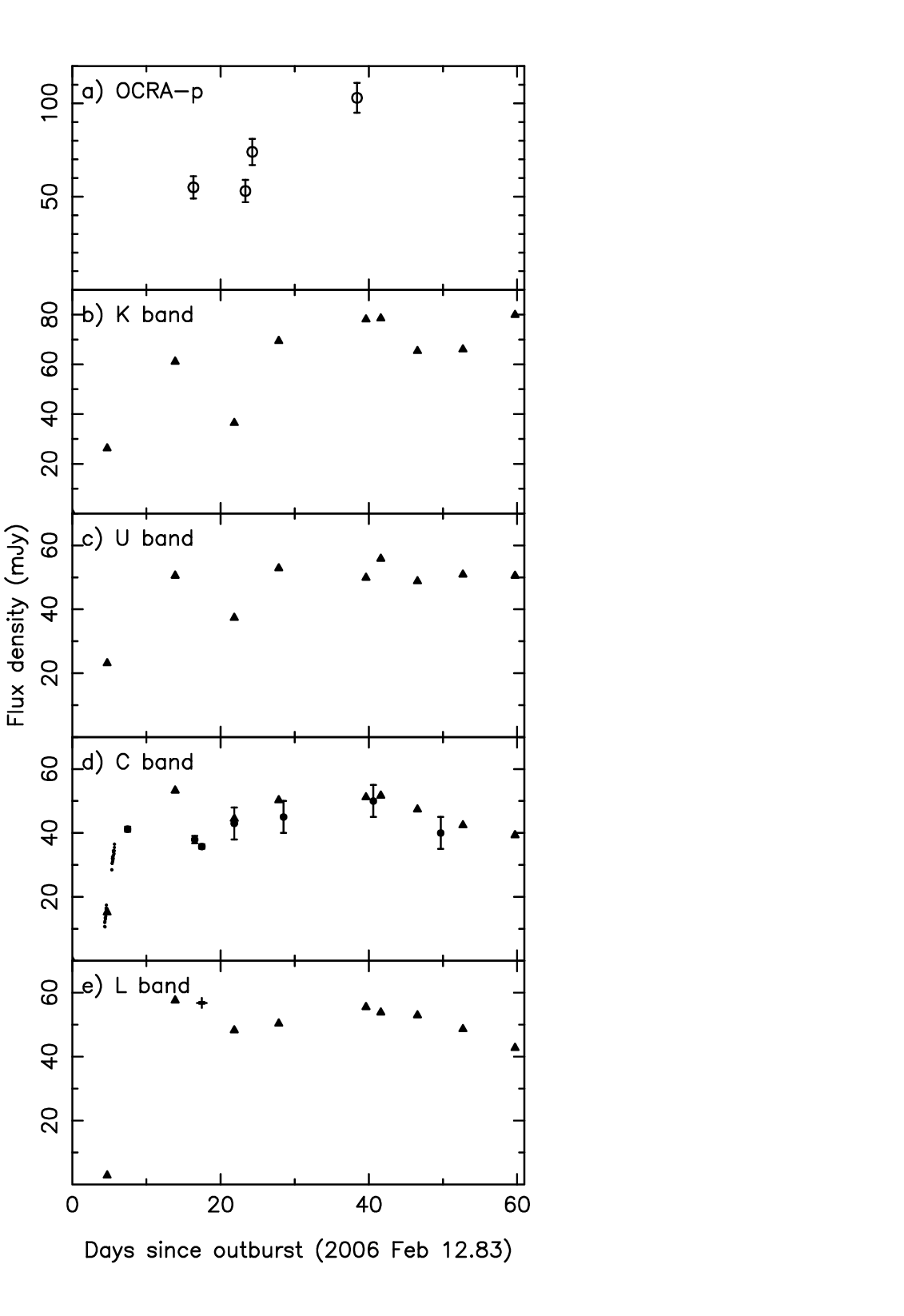}
\caption{Radio lightcurves at different frequencies. (a) 30~GHz with
  OCRA-p. (b) K~band -- 22.48~GHz with the VLA. (c) U~band --
  14.96~GHz with the VLA. (d) For comparison: C~band -- 4.88~GHz with
  the VLA, 6~GHz with MERLIN. (e) L~band -- 1.46~GHz with the VLA,
  1.39~GHz with the GMRT and 1.66~GHz with MERLIN. Symbols: VLA=filled
  triangles; MERLIN=filled circles (small ones during the initial rise
  -- error bars not included); GMRT=crosses; OCRA-p=open circles.
  Uncertainties are smaller than or comparable to the symbol size
  except where error bars indicate otherwise.}
\label{fig-LCs}
\end{figure}

The light curves in Figs.~\ref{fig-Cband} and \ref{fig-LCs} show the
variation in flux density in each of the four frequency bands observed
with the VLA and the one observed with OCRA-p. We have also included
observations from MERLIN and the GMRT with these bands, where the
spectral energy distribution (SED) indicates that they are comparable
despite the slightly different observation
frequency. Fig.~\ref{fig-Cband} shows particularly clearly the double
radio peak, and the difference from the 1985 data also plotted.

The spectral index at C~band was sufficiently flat (see
section~\ref{ssec-SEDs}) to allow us to plot all the MERLIN C~band
observations with the corresponding VLA C~band observations, even
though they were at somewhat different frequencies. The main features
of the light curves are as follows:

\begin{enumerate}
\item
An initial rapid rise, captured particularly well by the segmented
MERLIN C~band data. The rise in that band occurred at a rate of
18.5$\pm$0.5~mJy~day$^{-1}$ over days~4 and 5, peaking before
day~13.88 when a VLA C~band measurement shows the flux density as
53.29~mJy and declining to 37.93~mJy by day~16.29 (MERLIN C~band). On
day~7 the data show a consistent flux of $\sim$41$\pm$1 in each
30--min bin across the approximately 6.5~hr observation. Taking the
uncertainty as a limit on the change in flux over that time, we find
the rate of change had dropped below $\sim$3.8~mJy~day$^{-1}$.  The
lightcurves at other bands are consistent with a similar rise and fall
at all observed wavelengths.
\item
A second rise starting before day~21.85 and peaking before
day~39. Extrapolating both the second rise and subsequent decline
suggests a peak around day~38.0 at $\sim$63~mJy (C~band). The same
pattern is evident at all bands observed with the VLA, although at
K~band either the minimum between the two peaks was deeper or the
second rise began later. The peak flux density measured with OCRA-p at 30~GHz is
consistent with this peak.
\end{enumerate}

\subsection{Radio spectrum}
\label{ssec-SEDs}

\begin{figure}
\includegraphics[angle=0,  height=20cm]{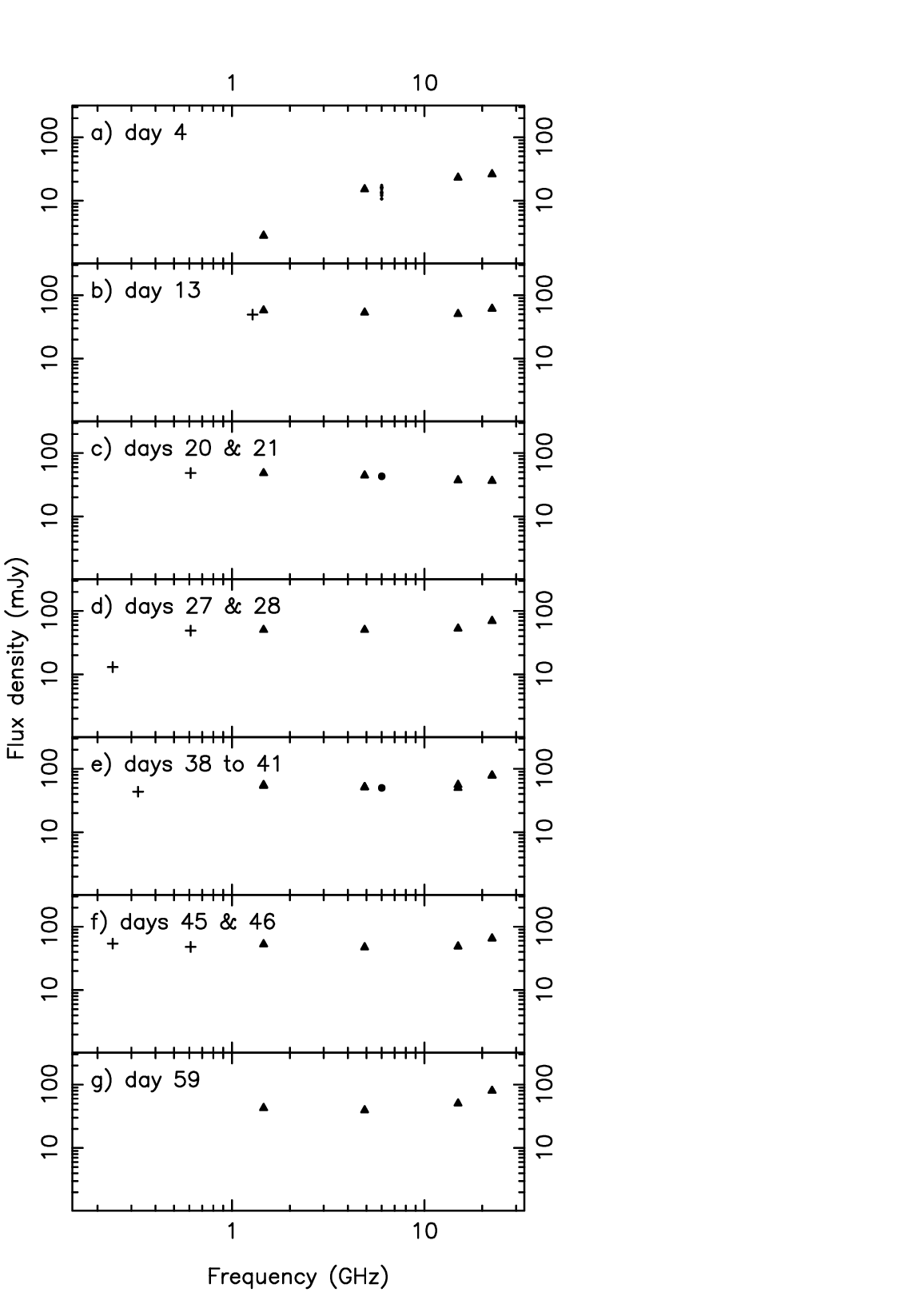}
\caption{Radio spectrum on various days as marked. Symbols and
  uncertainties as Fig~\ref{fig-LCs}.  Where available 240, 325 or
  610~MHz points (crosses) from \citet{Kantharia2007} are included,
  demonstrating the development of emission at lower frequencies.}
\label{fig-SEDs}
\end{figure}

The radio broad--band spectrum varies considerably over the course of
our observations, as seen in Fig.~\ref{fig-SEDs}. 
We can see immediately the effects of the rapid rise in the flux
density at 1.46~GHz, as the spectrum changes from positive spectral
index $\alpha$ (S$_\nu\propto\nu^\alpha$ convention) across the
observing bands on day~4 to having a minimum between 4.89 and
14.86~GHz on day~13 and a flat spectrum around day~21. By day~27 and
up to day~46 we see a return to the central minimum. The detailed
band--to--band spectral index values for the VLA observations are
given in Table~\ref{tab-alpha}, and are also apparent in the plots of
Fig.~\ref{fig-SEDs}.



\begin{table}
\begin{tabular}{lccc}
Day & \multicolumn{3}{c}{Band--to--band spectral index}\\
 & L to C & C to U & U to K\\
\hline
4.7 & 1.40$\pm$0.07 & 0.38$\pm$0.03  & 0.3$\pm$0.1\\
13.88 & $-$0.064$\pm$0.009 & $-$0.05$\pm$0.01 & 0.47$\pm$0.05\\
21.85 & $-$0.068$\pm$0.009 & $-$0.16$\pm$0.02 & 0.0$\pm$0.1\\
\hline
27.85 & 0.0$\pm$0.01 & 0.05$\pm$0.01 & 0.67$\pm$0.05\\
39.62 & $-$0.067$\pm$0.007 & $-$0.02$\pm$0.02 & 1.10$\pm$0.05\\
41.62 & $-$0.034$\pm$0.007 & 0.07$\pm$0.01 & 0.84$\pm$0.05\\
\hline
46.57 & $-$0.092$\pm$0.008 & 0.03$\pm$0.01 & 0.72$\pm$0.05\\
52.71 & $-$0.11$\pm$0.01 & 0.16$\pm$0.02 & 0.64$\pm$0.04\\
59.74 & $-$0.07$\pm$0.01 & 0.22$\pm$0.02 & 1.12$\pm$0.08\\ 
\hline
\end{tabular}
\caption{Simple band--to--band spectral index ($\alpha$ with flux
  S$_\nu\propto\nu^\alpha$ convention) for VLA observations on
  different days since optical discovery. The variable contribution of
  the thermal emission is particularly evident in the U to K band
  spectral index.}
\label{tab-alpha}
\end{table}

\subsection{Resolved emission at K~band (22.48~GHz)}
\label{ssec-Kband}

While none of the images at K band are sufficiently resolved to allow
us to identify structure within the emission, many have sufficient
resolution to allow us to determine the overall size of the emission
region following deconvolution. The results of a simple Gaussian
deconvolution on nine epochs of K~band data are given in
Table~\ref{tab-Kband}.

\begin{table}
\begin{tabular}{lccccc}
Days &  \multicolumn{3}{c}{Deconvolved extent} &\multicolumn{2}{c}{Fitted beam size}\\
& Major & Minor & PA & Major $\times$ Minor & PA\\
\hline
4   & 42$\pm$3 & 23$\pm$3 & 150$\pm$8 & 133$\times$83 & $-$24.84\\
13  & 27$\pm$4 & 23$\pm$3 &  20$\pm$4 & 149$\times$87 &    34.84\\
21  & 42$\pm$2 & 33$\pm$3 &  83$\pm$11 & 145$\times$95 &   34.06\\
\hline
27  & 43$\pm$2 & 28$\pm$4 &  97$\pm$8 & 148$\times$87 &    34.28\\
39  & 64$\pm$1 & 31$\pm$2 &  87$\pm$2 & 127$\times$82 &    20.98\\
41  & 65$\pm$1 & 27$\pm$4 &  92$\pm$2 & 131$\times$81 &    27.08\\
\hline
46  & 69$\pm$1 & 53$\pm$2 & 105$\pm$4 & 121$\times$80 &     7.44\\
52  & 87$\pm$1 & 45$\pm$2 &  92$\pm$2 & 121$\times$78 &     6.74\\
59  & 90$\pm$3 & 51$\pm$4 &  76$\pm$4 & 143$\times$84 &    33.83\\
\end{tabular}
\caption{Extent of VLA K~band deconvolved emission on different days
  since optical discovery. Deconvolved sizes (major and minor axes)
  are in milli--arcsec (mas), beam size (major and minor axes) in mas
  and position angles (PA) in degrees (positive
  north--through--east). All sizes are FWHM of two--dimensional
  Gaussians.}
\label{tab-Kband}
\end{table}

In all cases the beam is elongated, with the major axis being between
50\% and 70\% longer than the minor axis. Some care must be applied
when interpreting deconvolved extents. Where there are
$\sim$0$^\circ$\ or $\sim$180$^\circ$\ between the two position angle
(PA) values (extent and beam) then the deconvolution is highly
uncertain as we are attempting to resolve in the direction of worst
resolution. This problem affects the deconvolution for Days~4 and 13,
where the difference between the PA values is $\sim$175$^\circ$\ and
$\sim$14$^\circ$\ respectively, and we do not include these points in
our subsequent discussion. In all but two of the other cases the
deconvolved PA$\sim$90$^\circ$, consistent with imaging results from
the VLTI, EVN, MERLIN, VLBA and HST \citep{Sokoloski2008,
  Chesneau2008, OBrien2006, Rupen2008, Bode2007}.

\subsection{VLBA image at 1.667~GHz}

\begin{figure}
\includegraphics[angle=-90, width=8.5cm]{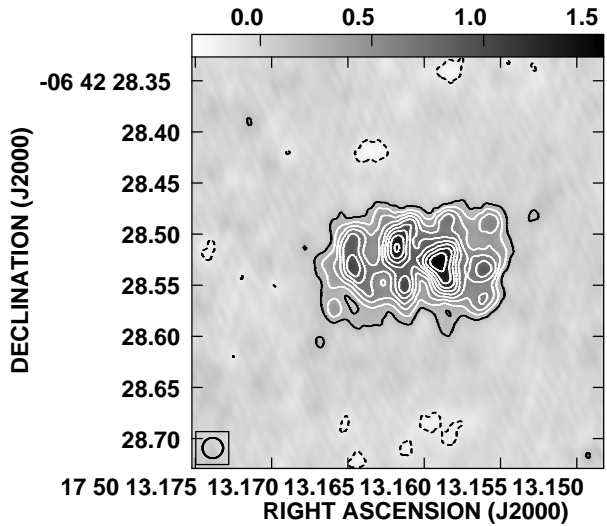}
\caption{VLBA image at 1.667~GHz taken on day~63. Beam size is
  indicated in the bottom left. Greyscale range is -0.302 to
  1.532~mJy~beam$^{-1}$. Contours are at intervals of 10\% of the
  peak of 1.532~mJy~beam$^{-1}$; negative contours indicated with
  broken lines.}
\label{fig-vlba}
\end{figure}

The image formed from VLBA data at 1.667~GHz on day~63
(Fig.~\ref{fig-vlba}) shows multiple components extended
east--west. The inner components are in the same place and of the same
size as the emission seen at 22.48~GHz with the VLA on day~59
(Table~\ref{tab-Kband}). Components extended east-west were seen in
earlier VLBA, EVN and MERLIN data \citep{OBrien2006, OBrien2008,
  Rupen2008, Sokoloski2008} so it is reasonable to assume these are
related to the more extended outer components seen here at day 63 and
hence would have also been present just four days earlier at the time
of the VLA imaging. 

We have attempted to estimate the spectral index of these
components. The eastern and brightest western components each have
flux densities in Fig.~\ref{fig-vlba} of $\sim$3.6~mJy. However this
image is heavily tapered, and these are really lower limits on the
flux density. Comparing this with a 1$\sigma$ upper limit for these
regions in the VLA 22~GHz data of day~59, we find $\alpha_{\rm LK} <
-0.3$, wholly consistent with these features being non--thermal. If
the actual flux density of the extended components is greater at
L~band, this value will be still more negative.

\section{Discussion}
\label{sec-discussion}


\subsection{Light curve development}
\label{ssec-LCdev}

In 1985 observations starting on day~18 were extrapolated back to an assumed
``turn-on'' of the emission on about day~14 \citep{Padin1985}.  It is
evident that the first peak seen in 2006 was before any observations
in 1985. However we note that the 2006 flux density on days~16 and 17 at 6~GHz
(Table~\ref{tab-obs}) is significantly higher than seen on day~18 in
1985. This implies either a fairly rapid drop of $\sim$12~mJy and
recovery of $\sim$20~mJy in $\sim$3~days, or that the initial peak was
completed more rapidly in 1985 (if present at all). In addition it is
notable that emission at lower frequencies was not seen in 1985, but
is evident after day~20 in 2006 \citep{Kantharia2007}.


It is not immediately clear what the cause of this early rise is, but
one possibility is that we are seeing emission from the reverse shock
traversing the ejecta at early times. However, below we suggest a
two--component model for the emitting regions.

A number of supernovae have shown variations in radio emission,
generally attributed to the ejecta interacting with density variations
in the surrounding medium \citep[e.g.][]{Weiler1991,Weiler1992}. This
may be a contributing factor in RS~Ophiuchi, where the density
variations may be in the wind from the red giant, although the
incomplete sampling of the radio light curve makes it difficult to
test this idea.


\subsection{Spectral development}
\label{ssec-SEDdev}

Our multi--band observations show that the SED varies over timescales
of a few days. The initial rapid rise at 1.46~GHz and 4.89~GHz is
consistent with the emergence of an expanding source from the densest
inner parts of an enveloping absorbing medium. This we identify with
the red giant wind ionised by the nova eruption, and providing a
source of free-free absorption. The rise between days~4 and 13 is at
least 6~mJy~day$^{-1}$, although as it is possible that the 5~GHz flux density
had peaked earlier this is a lower limit. By day~13 we are seeing a
mixed optically thin and thick thermal spectrum, and it is striking
that this is similar to the SED on days~38 to 46. Both these periods
are at, or shortly after, a radio peak, suggesting similar emission
details at these epochs. Around day~21, the brightness minimum between
the two peaks, the spectrum declines with frequency across the band,
indicating that the optically thin thermal emission dominates at this
time. Note also that \citet{Kantharia2007} have demonstrated a
non-thermal emission component dominates below 1.4~GHz from
day~20. Thus while a thermal spectrum dominates at high--frequencies,
a non--thermal spectrum dominates at low frequencies. Consequently in
most of the bands observed with the VLA a combined spectrum is seen.

We note that the best fitting supernova--style model (fading
non--thermal emission subject to variable absorption) of
\citet{Kantharia2007} also fits our early L--band data (up to and
including day 21). However it does not fit the higher frequency data
at these early times, predicting much brighter C--band emission that
turns on at earlier times than is observed. Furthermore, it clearly
cannot fit the subsequent (second) peak seen in our L--band and higher
frequencies. The data suggest at least two emitting components subject
to differing absorption and likely to be a mix of thermal and
non-thermal emission. The complexity of the geometry and emission
details precludes a detailed fit to the full dataset at this
stage.

\subsection{Ejecta morphology}

The spectral development can be further constrained by examining the
morphological development of the ejecta at different wavelengths.

Day~13 coincides with a simple, if one--sided, emission shell
\citep[VLBI images]{OBrien2006}. Day~21 sees the emergence of an
eastern component that also experiences absorption such that it
appears at 5~GHz before 1.46~GHz \citep{OBrien2006}.


The expansion apparent from the 22~GHz images (Table~\ref{tab-Kband})
is consistent with a power law of the form angular size $\theta_{\rm
K} = (2.1\pm0.1)t^{0.9\pm0.1}$~mas, where time $t$ in days has its origin on
day~0. 
Although we discard the size measurement for day~13 at 22~GHz
(section~\ref{ssec-Kband}), this expansion rate is consistent with the
size of the partial shell structure seen by \citet{OBrien2006} on that
day. If expansion proceeded at a constant rate at later times, this
also agrees with the E-W extent of the remnant imaged by the HST on
day~155 \citep{Bode2007} At a distance of 1.6~kpc
\citep{Hjellming1986} this equates to a projected velocity of
$\sim$4000~km~s$^{-1}$, constant within the errors (or equivalently
negligible deceleration over this period, within measurement
uncertainties). This 22~GHz emission is that from the inner shell seen
by \citet{OBrien2006}, as we discuss below. However the data are
inconsistent with a linear expansion from the time of ejection, as
this would require an angular extent of $\sim$9~mas on day~0. Thus
these VLA data imply a significant deceleration in the first 2~weeks
after optical discovery, consistent with results in the X-ray
waveband, related to deceleration of the forward shock by
\citet{Bode2006}.



 We can relate these measurements to resolved structure at
 1.667~GHz. The VLBA image at 1.667~GHz on day~63
 (Fig.~\ref{fig-vlba}) shows an extent of $\sim$200~mas
 east--west. This should have been resolved by the VLA on day~59. The
 fact it was not indicates that the more extended features have a
 spectrum that declines steeply with frequency, consistent with the
 outlying features being due to non-thermal emission. This is
 supported by the limit on the spectral index of $\alpha_{\rm LK} <
 -0.3$. Thus we are able to demonstrate that the initial shell seen by
 \citet{OBrien2006} has a contribution from thermal component (as it
 is brightest at the higher frequencies), while the more extended
 emission is entirely non-thermal in nature. This is consistent with
 the findings of \citet{Sokoloski2008} and strongly confirms the
 conclusion of \citet{Taylor1989} with regard to the 1985 outburst.

This is also comparable to the results of \citet{Crocker2001} for for
the symbiotic star CH~Cyg during ejection episodes. They found both
non-thermal extended jet-like structures and a thermally-dominated
central source.

\subsection{Ejecta mass}
\label{ssec-mass}


We make use of the formulation of \citet{Mezger1967} to estimate the
total ejecta mass.  We use the emission on day~4.7, as at later dates
it is not clear what geometry we should use -- at this early time the
actual distribution of matter in the emission region is most likely to
be reasonably represented by a filled sphere. While there are more
physically realistic models of the emission geometry possible, the
uncertainties make such considerations relatively unimportant. We make
a number of assumptions: that the emission is optically thin, that
50\% is thermal, that we can extrapolate back from the t$^{0.9}$
dependence of the 22.48~GHz angular extent to day~4.7, and finally
that the electron temperature is 10$^4$~K. This gives an angular size
of 8.5$\pm$0.2~mas (formal uncertainty) and a thermal flux density of
13.1$\pm$0.1~mJy. From this we find a mass of
4$\pm$2$\times10^{-7}$~M$_\odot$.

The dependence on electron temperature is very weak (a power of 0.175)
so the last assumption has little impact. The mass depends on the
square root of the flux density, so a factor of two larger or smaller only
changes the estimate by a factor of $\sqrt2$. The strongest
dependencies are on angular size and distance. A 30\% larger angular
size leads to a 50\% larger mass estimate, but such a large size is
well outside any reasonable estimate given what we see at later
dates. The dominant uncertainty is in the distance, at 1.6$\pm$0.3~kpc
\citep{Hjellming1986}. This gives most of the 50\% fractional
uncertainty in the mass estimate. From this we can see that the radio
emission at the earliest times is most consistent with the ejecta mass
associated by modelling \citep{Yaron2005} with a WD closer to
1.4~M$_\odot$ than the lower values considered, excluding cases with
recurrence intervals of less than a year. A higher ejecta mass by a
factor of 10 could be made consistent with a lower WD mass of
1.25~M$_\odot$.

\section{Conclusions}
\label{sec-conclusions}

We present 63 days of radio observations of the 2006 outburst of
RS~Oph with MERLIN and the VLA, with reference to additional
observations with the VLBA, GMRT, Effelsberg and OCRA-p. The
observations demonstrate that the emission had a mixture of thermal
and non-thermal components beginning no later than 13 days after the
outburst. An early radio peak before day 13 is seen for the first
time. The early radio rise (up to day 5.5) is consistent with the
transition from the ``free--expansion'' phase noted by
\citet{Bode2006} and \citet{Das2006}.

At the minimum between the first and second peaks the flux density
declines slightly across the frequency bands, consistent with mixed
thermal and non--thermal emission, but analysis of our imaging results
clearly confirm the non-thermal nature of the most extended components
in the VLBI imaging of \citet{OBrien2006}. At and after the second
peak the spectral behaviour returns to a mixture of thermal and
non-thermal emission. Comparison of the 22~GHz VLA observation on
day~59 with the 1.667~GHz VLBA observation on day 63 indicates that
the emerging east and west structures are non-thermal in nature with a
spectral index $\alpha < \sim-0.3$.  In addition size estimates
between days 27 and 59 indicate a decelerating expansion. We find an
ejecta mass of 4$\pm$2$\times10^{-7}$~M$_\odot$, but this is probably
a lower limit due to the uncertainty surrounding the overlying
absorption component and the assumptions made in making the
estimate. Comparison with the modelling of \citet{Yaron2005} imply a
WD mass around 1.4~M$_\odot$; a higher ejecta mass would imply a lower
WD mass.

This radio development is consistent with two components -- an
expanding, decelerating shell seen in both non--thermal and thermal
emission plus bipolar ejecta generating non--thermal emission, also
seen by \citet{Sokoloski2008}. The mechanism for generating such
bipolar structures in this system requires further study, and may tell
us more about the mass--loss and mass--transfer processes in recurrent
novae. However it is consistent with the interpretation by
\citet{Taylor1989} of the only equivalent data for the 1985 outburst,
and also similar to the interpretation of radio jets during an
outburst of the symbiotic star CH~Cygni \citep{Crocker2001}. This is
evidence of another non--thermal WD jet from a symbiotic system during
an ejection episode.

\section*{Acknowledgments}

\noindent MFB was supported by a PPARC/STFC Senior Fellowship.  MTR
was funded by the University of Central Lancashire. The Toru{\'n}
group acknowledge support via the TCfA grant KBN~5~P03D~024~21. MERLIN
is a National Facility operated by the University of Manchester at
Jodrell Bank Observatory on behalf of the Science \&\ Technology
Facilities council. The National Radio Astronomy Observatory is a
facility of the National Science Foundation operated under cooperative
agreement by Associated Universities, Inc.

\bsp

\label{lastpage}
\end{document}